\documentclass[namedreferences]{solarphysics}

\usepackage[hyperref,optionalrh]{spr-sola-addons} 
\usepackage{graphicx}        
\usepackage{color}           
\usepackage{breakurl}        

\newcommand{\aap}{    {\it Astron. Astrophys.}}

\newcommand{\apj}{    {\it Astrophys. J.}}
\newcommand{\apjl}{   {\it Astrophys. J. Lett.}}

\newcommand{\cjaa}{   {\it Chin. J. Astron. Astrophys.}}

\newcommand{\jgr}{    {\it J. Geophys. Res.}}

\newcommand{\solphys}{{\it Solar Phys.}}

\newcommand{\ssr}{    {\it Space Sci. Rev.}}

\begin{document}

\begin{article}

\begin{opening}

\title{Filament Activation in Response to Magnetic
Flux Emergence and Cancellation in Filament Channels}

\author{Ting~\surname{Li}$^{1}$\sep
        Jun~\surname{Zhang}$^{1}$\sep
        Haisheng~\surname{Ji}$^{2}$
       }
\runningauthor{Li et al.}
\runningtitle{Filament activation}

   \institute{$^{1}$ Key Laboratory of Solar Activity, National
Astronomical Observatories, Chinese Academy of Sciences, Beijing
100012, China;
                     email: \url{liting@nao.cas.cn}\\
              $^{2}$ Key Laboratory for Dark Matter and Space Science,
Purple Mountain Observatory, CAS, Nanjing 210008, China
                      \\
             }

\begin{abstract}
We make a comparative analysis for two filaments that showed quite
different activation in response to the flux emergence within the
filament channels. The observations from the \emph{Solar Dynamics
Observatory} (\emph{SDO}) and Global Oscillation Network Group
(GONG) are carried out to analyze the two filaments on 2013 August
17-20 and September 29. The first event showed that the main body of
the filament was separated into two parts when an active region (AR)
emerged with a maximum magnetic flux of about 6.4$\times$10$^{21}$
Mx underlying the filament. The close neighborhood and common
direction of the bright threads in the filament and the open AR fan
loops suggest similar magnetic connectivity of these two flux
systems. The equilibrium of the filament was not destroyed within 3
days after the start of the emergence of the AR. To our knowledge,
similar observations have never been reported before. In the second
event, the emerging flux occurred nearby a barb of the filament with
a maximum magnetic flux of 4.2$\times$10$^{20}$ Mx, about one order
of magnitude less than that of the first event. The emerging flux
drove the convergence of two patches of parasitic polarity in the
vicinity of the barb, and resulted in cancellation between the
parasitic polarity and nearby network fields. About 20 hours after
the onset of the emergence, the filament was entirely erupted. Our
findings imply that the location of emerging flux within the
filament channel is probably crucial to filament evolution. If the
flux emergence appears nearby the barbs, flux cancellation of
emerging flux with the filament magnetic fields is prone to occur,
which probably causes the filament eruption. The comparison of the
two events shows that the emergence of an entire AR may still not be
enough to disrupt the stability of a filament system and the actual
eruption does occur only after the flux cancellation sets in.
\end{abstract}
\keywords{Prominences, Formation and Evolution; Prominences,
Dynamics; Magnetic fields, Photosphere; Coronal Mass Ejections,
Initiation and Propagation}
\end{opening}

\section{Introduction}
     \label{}

The relationship between filament activation and the evolution of
its surrounding magnetic fields is of great importance for
understanding the triggering mechanisms of the activation (Wang et
al. 1996; Schmieder et al. 2008; Kusano et al. 2012). Since the
associated magnetic changes usually occur earlier than the filament
activation (Jiang et al. 2007; Sterling et al. 2007; Xu et al.
2008), the evolution of magnetic fields in the filament environment
can serve as an early signature for the occurrence of filament
eruptions and the associated coronal mass ejections (CMEs).

The slow$-$rise motion of the filament before the rapid acceleration
is frequently triggered by magnetic flux emergence or cancellation
(Forbes \& Isenberg 1991; Chen et al. 2009; Shen et al. 2011; Bi et
al. 2013). The slow magnetic reconnection in the process of flux
emergence and cancellation gradually changes the equilibrium state
of a filament and usually results in the slow ascent of the filament
(Chen \& Shibata 2000; Lin et al. 2001; Zhang \& Wang 2001;
Nagashima et al. 2007). The newly emerging flux with a favorable
orientation for reconnection with the preexisting flux is more
likely to lead to an eruption than the one with an unfavorable
orientation (Feynman \& Martin 1995; Galsgaard et al. 2007). Jing et
al. (2004) found that 54 events (68\%) of 80 filament eruptions were
accompanied by new flux emergence adjacent to eruptive filaments
that usually appeared within 15 hr before filament eruptions.

Regarding the magnetic field topology of filaments, several models
have been proposed so far (Mackay et al. 2010). In the empirical
`wire model,' the magnetic field lines of a filament are highly
sheared along the magnetic polarity inversion line (PIL) and do not
contain dips (Martin \& Echols 1994; Lin et al. 2008). The magnetic
fields supporting the barbs (plasma condensations) are vertical,
connecting the filament body to parasitic (minority) polarity
elements in the photosphere (Martin \& Echols 1994; Lin et al.
2005). The parasitic polarity is defined as the polarity opposite to
the dominant polarity of the network fields on the same side of the
filament. In sheared arcade model (Antiochos et al. 1994; DeVore \&
Antiochos 2000) and flux rope model (van Ballegooijen \& Martens
1989; Amari et al. 1999), the dense filament plasma lies in the dips
of the magnetic fields. Aulanier et al. (1998, 1999) have developed
3D magnetic models of filaments and showed that the filaments can be
represented by helical flux ropes overlying the PIL. The plasma of
barbs is confined by the magnetic dips and the endpoints of barbs in
the photosphere are located at secondary PIL separating the
parasitic polarity from the surrounding dominant flux (Dud{\'{\i}}k
et al. 2008).

In this work, we make a comparative analysis for two events focusing
on the relationship between filament activation and flux emergence
with the observations from the \emph{Solar Dynamics Observatory}
(\emph{SDO}; Pesnell et al. 2012) and Global Oscillation Network
Group (GONG). The first event showed that the main body of the
filament was separated into two parts when an active region (AR)
emerged within the filament channel. To our knowledge, the
observations of filament separation due to an emerging AR underlying
a filament have never been reported before. In the second event, the
emerging flux happened to appear in the vicinity of a barb, and
subsequently resulted in a filament eruption. This rarely reported
event gives us a good opportunity to study the interaction of the
emerging flux with the magnetic structures of a barb and the
physical process of how the evolution of a barb destabilizes the
main body of the corresponding filament.

\section{Observations and Data Analysis} 
      \label{}

The Atmospheric Imaging Assembly (AIA; Lemen et al. 2012) onboard
the \emph{SDO} takes full-disk images in 10 (E)UV channels at
1.5$^{\prime\prime}$ resolution and high cadence of 12 s. The
observations of 304 {\AA}, 171 {\AA} and 193 {\AA} are used in this
study. The channel of 304 {\AA} (He II) corresponds to a temperature
of about 0.05 MK, 171 {\AA} (Fe IX) at 0.6 MK, and the 193 {\AA}
channel (Fe XII) is at 1.5 MK (with a hot contribution of Fe XXIV at
20 MK, Ca XVII at 6.0 MK and cooler O V at 0.2 MK; O'Dwyer et al.
2010; Del Zanna 2013). In order to investigate the interaction of
emerging flux with the ambient magnetic structures, we also use the
full-disk line-of-sight magnetic field data from the Helioseismic
and Magnetic Imager (HMI; Scherrer et al. 2012) onboard \emph{SDO},
with a cadence of $\sim$ 3 min and a sampling of
0.5$^{\prime\prime}$ pixel$^{-1}$. NSO$-$GONG H$\alpha$ data are
used to investigate the chromospheric configuration of the filament.
GONG collects H$\alpha$ data at seven sites with a spatial
resolution of 1.0$^{\prime\prime}$ pixel$^{-1}$ and a cadence of
around 1 min (Harvey et al. 2011).

\section{Results}
\subsection{The First Event (2013 August 17$-$20)}

The first event analyzed here occurred on 2013 August 17$-$20. A
large filament lies across the entire south hemisphere (Figure 1(d);
see Animation 0817-304). It has a length of about 992 Mm, and is
located between AR 11818 and AR 11820. Starting from 04:00 UT on
August 17, new magnetic flux emerged in the filament channel. The
two patches with opposite polarities gradually grew up and evolved
into AR 11824 (Figures 1(a)-(c)). Seen from AIA 304 {\AA}
observations, the emergence of magnetic flux is associated with EUV
intensity enhancement and no barbs are observed nearby the region of
flux emergence (Figures 1(e) and (f)). From 10:00 UT on August 17,
the main body of the filament was gradually divided into two parts
(Figures 1(d)$-$(f)). Due to the interaction between the AR and the
filament, the south end of the north filament is located in the
vicinity of the emerged positive polarity fields and the north end
of the south filament is located nearby the emerged negative
polarity fields (Figure 1(c)).
\begin{figure}
   \centerline{\includegraphics[width=1.0\textwidth,clip=]{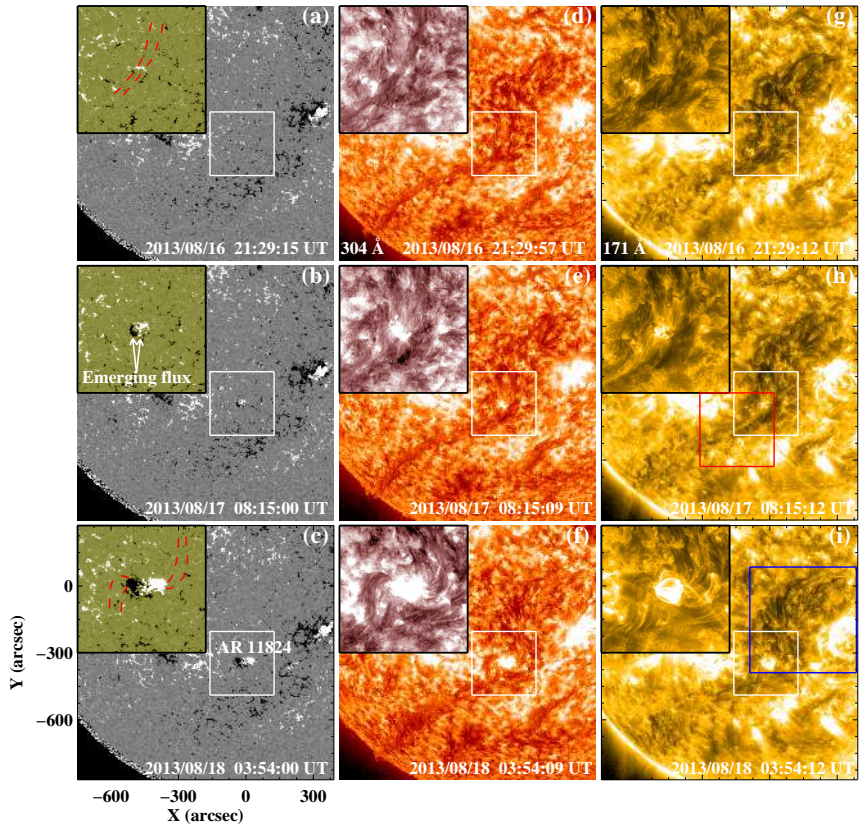}
              }
   \caption{{\emph{SDO}}$/$HMI line-of-sight magnetograms showing the emerging magnetic flux within
the filament channel (panels (a)$-$(c)), AIA 304 {\AA} and 171 {\AA}
images showing the evolution of the filament (panels (d)$-$(i); see
Animation 0817-304) during 2013 August 16$-$18. The upper-left
corner in each panel shows the enlarged image in the white square.
The red dashed lines in panels (a) and (c) denote the contours of
the filament. The red square in panel (h) denotes the field of view
(FOV) of Figures 3(a)$-$(f). The blue square in panel (i) denotes
the FOV of Figures 3(g)$-$(l).}
              \label{Fig1}%
    \end{figure}

\begin{figure}
   \centerline{\includegraphics[width=1.0\textwidth,clip=]{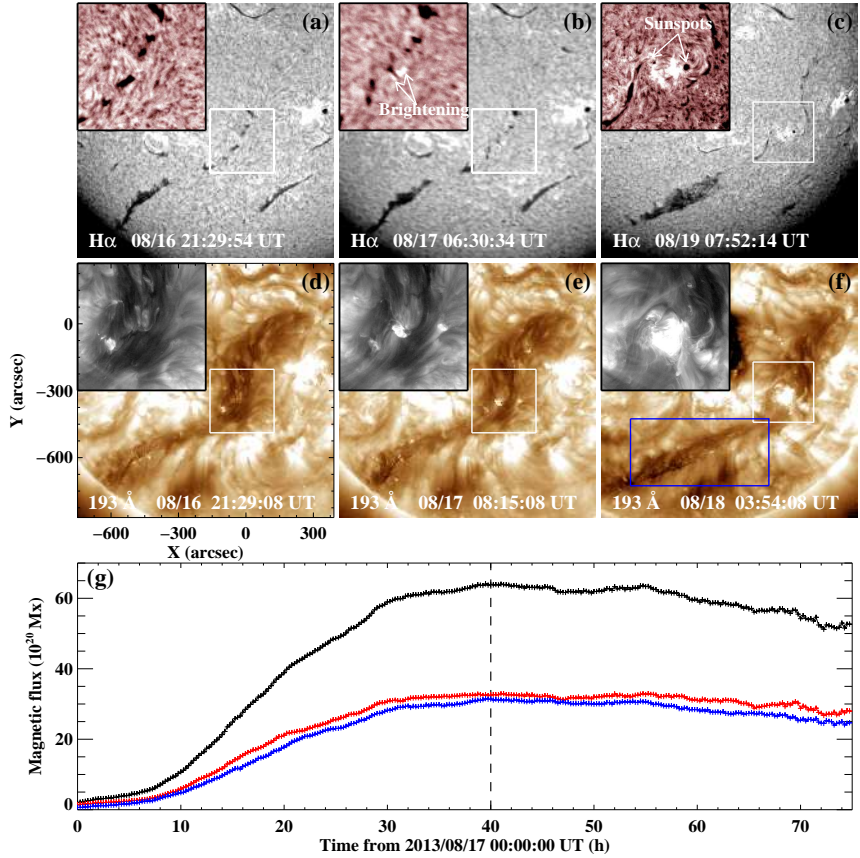}
              }
   \caption{Appearance of the first event in GONG H$\alpha$ (panels
(a)$-$(c)) and AIA 193 {\AA} (panels (d)$-$(f)) and temporal
variations of the emerging magnetic flux in HMI magnetograms (panel
(g)). The upper-left corner in each panel shows the enlarged image
in the white square. The blue rectangular in panel (f) denotes the
FOV of EUV and H$\alpha$ images in Figure 5. The black profile in
panel (g) denotes the evolution of total magnetic flux and the red
and blue profiles denote the unsigned positive and negative magnetic
flux. Dashed line in panel (g) denotes the time when the total
unsigned magnetic flux reached its maximum.}
              \label{Fig2}%
    \end{figure}

\begin{figure}
   \centerline{\includegraphics[width=1.0\textwidth,clip=]{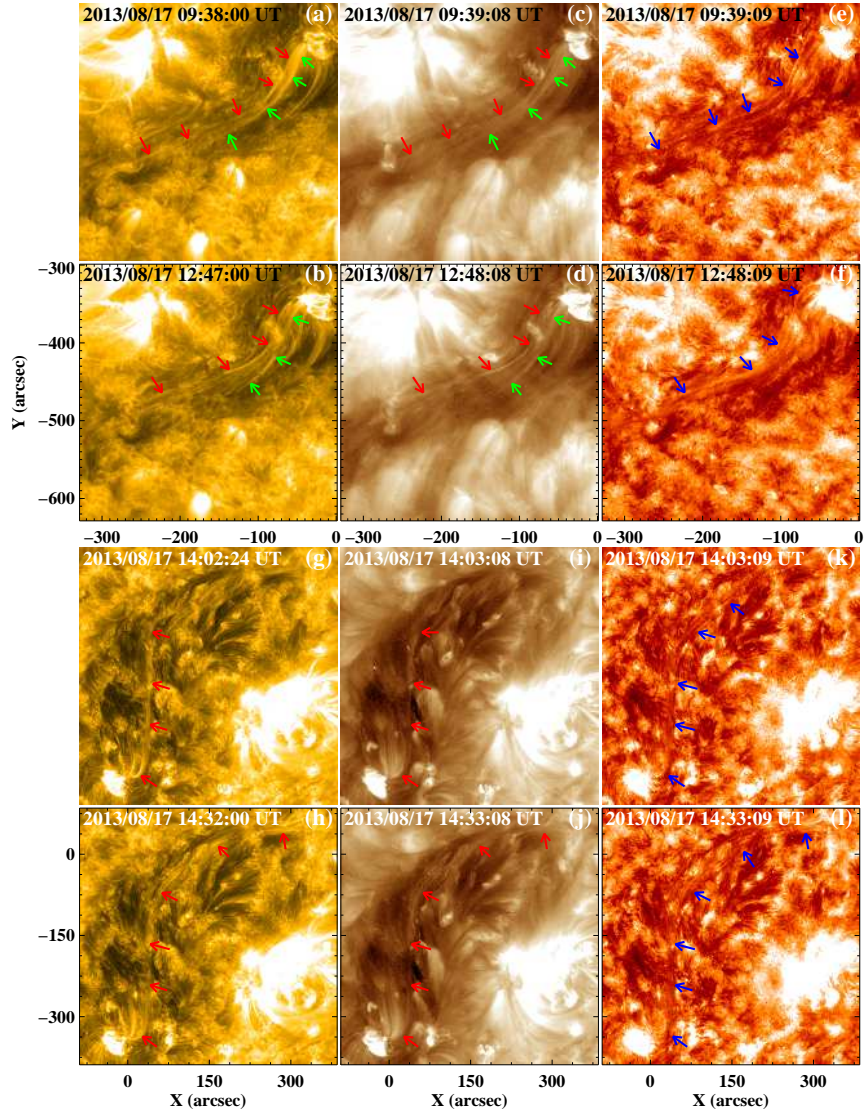}
              }
   \caption{{\emph{SDO}}$/$AIA 171, 193 and 304 {\AA} images showing the fine-scale structures of
the southern (panels (a)$-$(f)) and northern (panels (g)$-$(l))
filaments. The FOV of panels (a)$-$(f) is the red square in Figure
1(h), and the FOV of panels (g)$-$(l) corresponds to the blue square
in Figure 1(i). The arrows point to the bright threads in the
filament.}
              \label{Fig3}%
    \end{figure}

\begin{figure}
   \centerline{\includegraphics[width=1.0\textwidth,clip=]{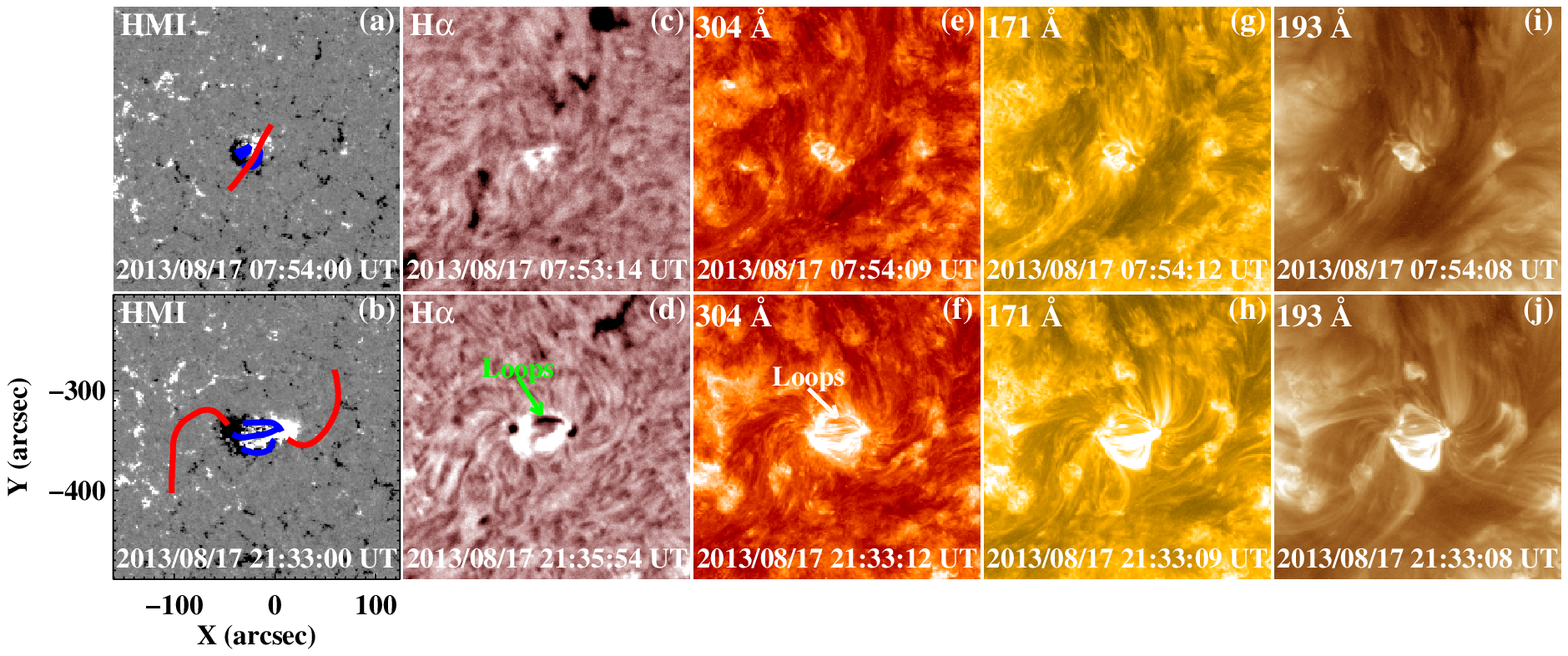}
              }
   \caption{Appearance of the filament and the closed loops of the AR in
HMI magnetograms, GONG H$\alpha$, AIA 304, 171 and 193 {\AA} images.
Red solid curves in panels (a) and (b) denote the axes of the
filament as seen in 304 {\AA} images and blue solid curves represent
the closed loops of the AR. The FOV of these images is denoted by
the white square in Figures 1 and 2.}
              \label{Fig4}%
    \end{figure}

In H$\alpha$ images (Figures 2(a)-(b)), the north part of the
filament is much fainter than that in EUV wavelength (Figures
1(d)-(f)). Nearby the emerging AR, the filament in H$\alpha$ is
fragmented and the connectivity could not be clearly observed. The
difference in H$\alpha$ and EUV images can be ascribed to more
abundant Lyman continuum absorption in EUV than in H$\alpha$ and the
effect of the coronal emissivity blocking due to the prominence void
(Wang et al. 1998; Heinzel et al. 2001, 2008; Schmieder et al. 2003;
Anzer \& Heinzel 2005; Li \& Zhang 2013a). Only the low and denser
parts of the filament may be visible in H$\alpha$ on the disk
(Heinzel et al. 2001). The appearances of the filament at 171 {\AA}
and 193 {\AA} are shown in Figures 1(g)$-$(i) and 2(d)$-$(f). The
171 {\AA} band can contain bright threads along the filament channel
due to the Fe IX line being in emission in the high-temperature
transition region (Parenti et al. 2012). The EUV filament channel is
more extended than the filament because of the local relative
absence of plasma with coronal densities and temperatures (i.e., a
`void'; Heinzel et al. 2008).

\begin{figure}
   \centerline{\includegraphics[width=1.0\textwidth,clip=]{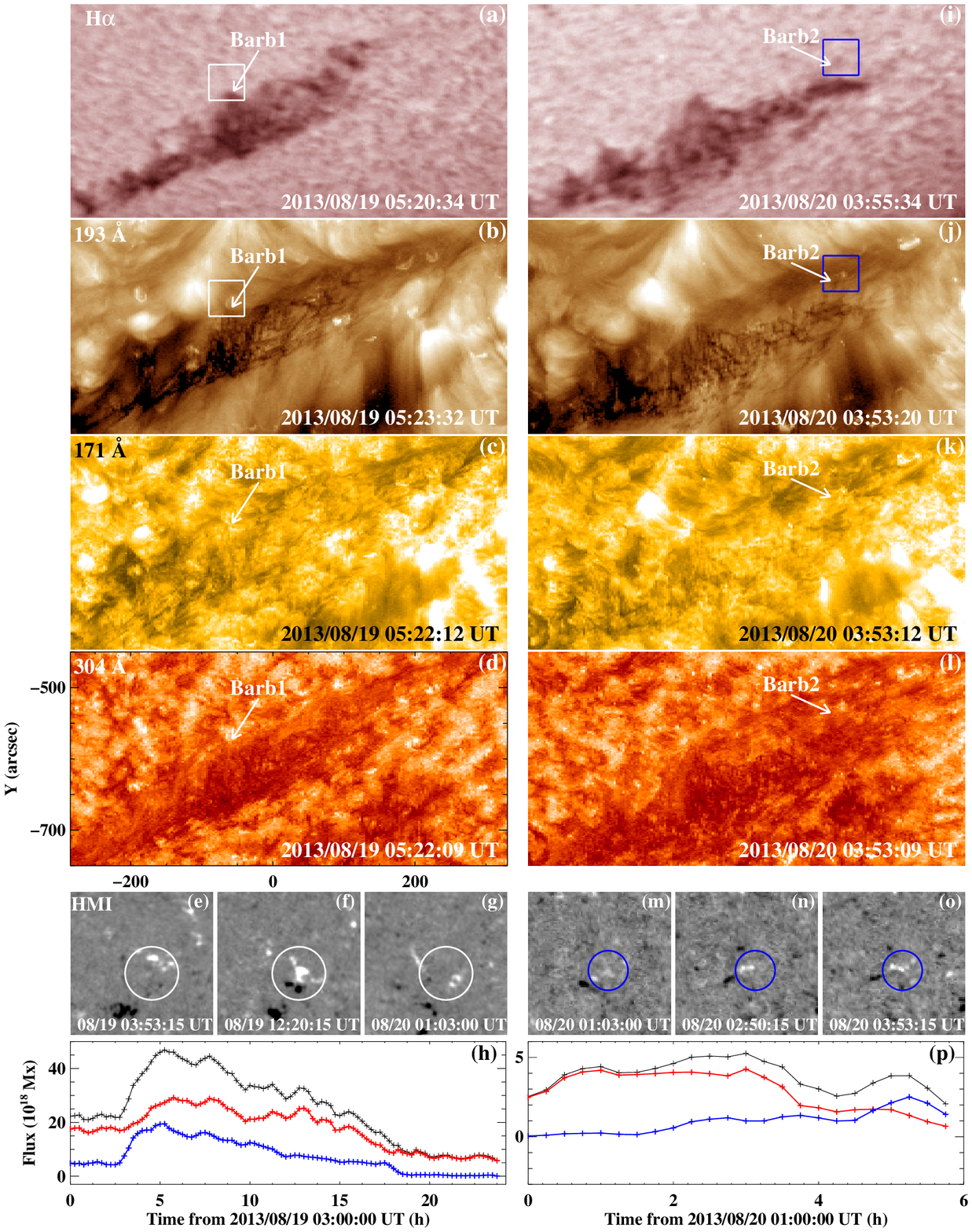}
              }
   \caption{GONG H$\alpha$ images, AIA 193, 171, 304 {\AA} images, HMI
magnetograms and temporal variations of the magnetic flux showing
two barbs of the southern filament and the evolution of magnetic
fields. The FOV of GONG H$\alpha$ and EUV images is represented by
the blue rectangular in panel (f) of Figure 2. The white square in
panels (a)-(b) denotes the FOV of panels (e)$-$(g) and the blue
square in panels (i)-(j) denotes the FOV of panels (m)$-$(o). White
circles outline the flux emergence and cancellation nearby the
Barb1, and blue circles outline the flux emergence nearby the Barb2.
The meanings of the profiles in panels (h) and (p) are the same as
those in Figure 2(g).}
              \label{Fig5}%
    \end{figure}
The magnetic flux gradually emerged underlying the filament and the
intensity enhancement in H$\alpha$ was simultaneously observed
(Figures 2(b)-(c)). Despite the emergence of the AR, the filament
was stable and did not erupt during August 18-19 (Figures 2(c) and
(f)). The H$\alpha$ filament is clearly visible in close vicinity of
the southeast and northwest parts of AR 11824. The leading and
following sunspots are formed at the emerging region (Figure 2(c)).
The leading sunspot has positive polarity fields and the following
one has negative polarity fields (Figure 1(c)).

The temporal variations of the emerging magnetic flux in HMI
magnetograms were measured and are displayed in Figure 2(g). We
derotated all the magnetograms differentially to a reference time
(2013 August 17 00:00 UT). The area within which we calculated the
magnetic flux is changed according to the expansion of the emerging
AR to ensure that the area has an appropriate size (Yang et al.
2012). Due to the expansion of the emerging flux during 3 days, a
pre-selected fixed box could not exactly contain the AR. If the box
is larger, other magnetic structures are also included. If the box
is smaller, the AR at the late stage is beyond the area of the box.
After selection, the pixels with unsigned magnetic fields weaker
than 10.2 Mx cm$^{-2}$ (noise level determined by Liu et al. 2012)
are eliminated. Moreover, the actual magnetic flux is approximately
$\Phi$/$\cos$($\alpha$$_{0}$)$\times$$\cos$($\alpha$) due to the
projection effects, where $\alpha$$_{0}$ is the heliocentric angle
at the reference time, $\alpha$ is the heliocentric angle at the
observation time and $\Phi$ is the observed magnetic flux. Here, we
used the equation to eliminate projection effects due to
differential rotation during construction of magnetic flux plots.

The temporal profiles of unsigned positive and negative magnetic
flux show consistent trend. The emergence lasted for about 36 hr and
the total magnetic flux increased to 6.4$\times$10$^{21}$ Mx at
16:00 UT on August 18. This is a smaller AR by comparing its
unsigned magnetic flux with typical unsigned fluxes of ARs in Warren
et al. (2012). The maximum values of positive and negative magnetic
flux are nearly equal, about 3.2$\times$10$^{21}$ Mx. The average
flux emergence rate is approximately 2.3$\times$10$^{16}$ Mx
s$^{-1}$. Afterwards, the positive and negative magnetic flux
gradually decreased and the morphologies of the north and south
filaments remained steady. At 03:00 UT on Aug 20, the total unsigned
magnetic flux decreased by 26\% within 35 hr.

At 09:09 UT on Aug 17, about 5 hr after the start of the flux
emergence, bright fine-scale threads could be clearly observed at
the southeast of the AR, with its end anchoring in the negative
polarity fields (Figures 3(a)-(f); Figure 1(b)). These bright
threads extended from the AR to the southeast within the filament
channel. The 171 {\AA} band best shows the bright threads that are
fainter at 193 and 304 {\AA}. Similar bright threads could be
observed in the northern part of the filament channel (Figures
3(g)-(j)). The ends of the threads are located at the positive
polarity fields (see Figure 1(c)). The 304 {\AA} observations also
show dark fine-scale threads of the filament connecting the AR with
the north end of the filament (Figures 3(k)-(l)). Accompanying the
interaction of the AR with the filament, the previous long filament
was divided into two parts, with the ends located nearby the two
opposite polarity fields of the AR. The long threads in the filament
channel usually could be observed in the passbands of 304 {\AA}, 171
{\AA} and 193 {\AA}. Thus the observed bright structures may contain
both the threads within the filament itself and the skewed fan loops
with one footpoint rooting in one polarity of the AR.

Moreover, the bright and dark closed loops appeared in the emerging
AR (Figure 4). The emerging AR showed a clockwise rotation and the
filament was cut into two parts, which formed two `S' shaped
structures. The closed loops were located between the two `S' shaped
structures (new north and south filaments). With the evolution of
the AR, the south leg of the north filament and the north leg of the
south filament were becoming more curved around the AR.

From about 05:30 UT on August 19, other small-scale magnetic flux
(Figures 5(e)-(f)) emerged nearby the Barb1 of the south filament.
Meanwhile, the EUV intensity enhancement at 304, 171 and 193 {\AA}
was observed near the Barb1 and the brightenings in H$\alpha$ could
not be clearly observed. The temporal variation of the emerging
magnetic flux is shown in panel (h). At 08:00 UT, the unsigned
magnetic flux increased to its maximum of about 4.7$\times$10$^{19}$
Mx. Then the emerged flux gradually canceled with nearby magnetic
fields and the unsigned magnetic flux decreased to
5.9$\times$10$^{18}$ Mx at 01:00 UT on August 20 (panels (g)-(h)).
Another magnetic flux emerged nearby the Barb2 from 01:00 UT on
August 20 (panels (m)-(p)). The emerging magnetic fields are weak
and have an unsigned magnetic flux of 5.0$\times$10$^{18}$ Mx. These
small-scale magnetic fields seemed to disturb the stability of the
southern filament. At about 04:00 UT on August 20, the south
filament started to rise up slowly and erupted partially. The
filament eruption caused a CME with an average velocity of 490 km
s$^{-1}$ in the FOV of Large Angle and Spectrometric Coronagraph
(LASCO; Brueckner et al. 1995) C2. The filament eruption was not
accompanied by any X-ray flare.

\subsection{The Second Event (2013 September 29)}

In the second event on 2013 September 29, the lateral barb of the
filament could be clearly identified at AIA 304, 171 and 193 {\AA}
and H$\alpha$ observations (Figure 6; see Animation 0929-304). In
H$\alpha$ observations, the filament is composed of two strands
intertwined with each other (Figure 6(d), similar to the
observations of Liu et al. (2008)). From about 00:00 UT on September
29, the magnetic flux emerged in the vicinity of the barb within the
filament channel (Figures 6(a)-(c)). The emergence of magnetic flux
is associated with EUV intensity enhancement at AIA 171 and 193
{\AA} (Figures 6(k) and (n)), which indicates coronal emission
rather than absorption by the filament material (Anzer \& Heinzel
2005). The brightening could not be clearly observed in H$\alpha$
and 304 {\AA} (Figures 6(e) and (h)). After the emergence, a
destabilization of the filament occurred near the barb and resulted
in the eruption of the filament (H$\alpha$ and EUV images in right
column of Figure 6). The entire eruptive filament could be observed
in EUV passbands, however, only the north part of the filament
appears in H$\alpha$ during the eruption (panel (f)).

    \begin{figure}
   \centerline{\includegraphics[width=1.17\textwidth,clip=]{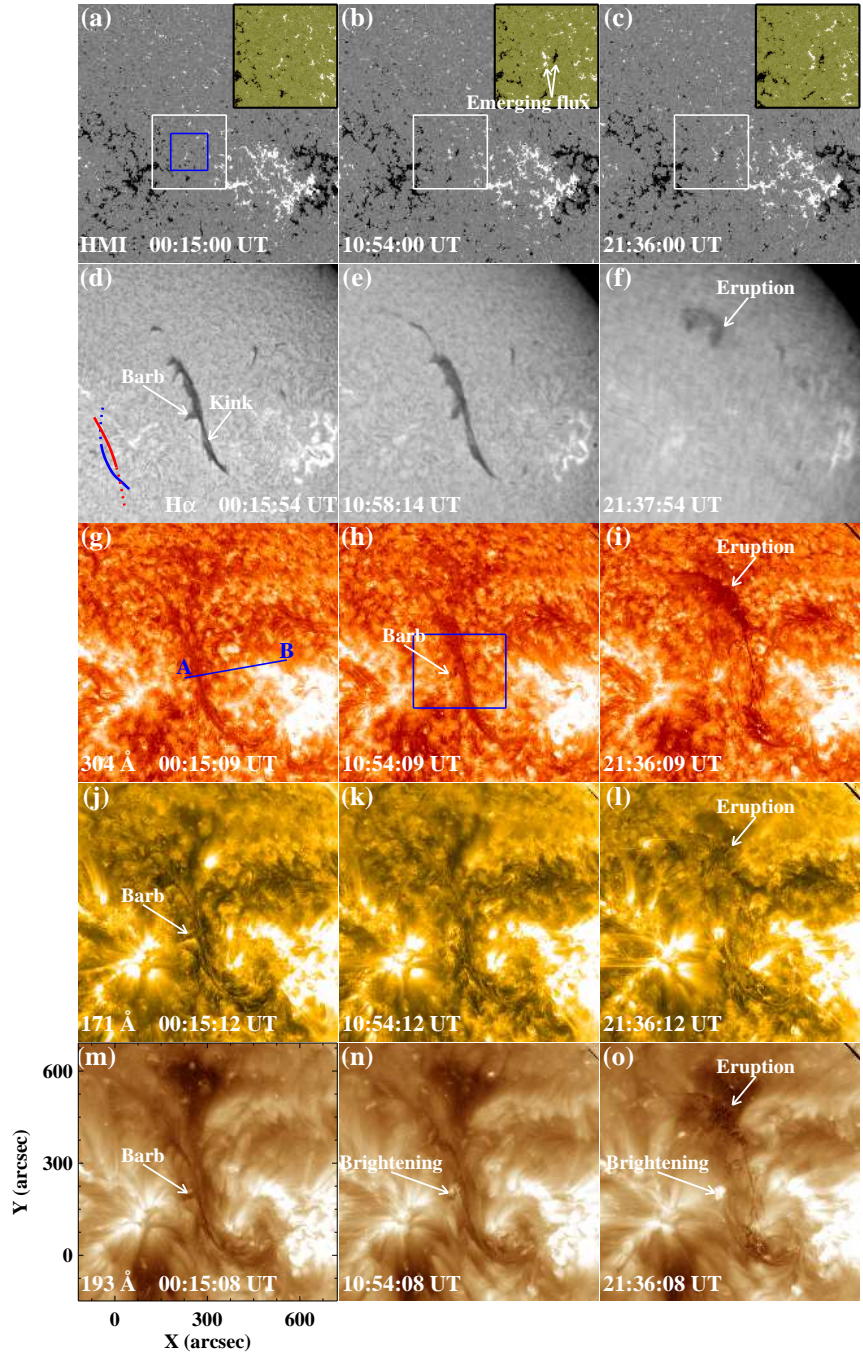}
              }
   \caption{{\emph{SDO}}$/$HMI magnetograms, GONG H$\alpha$, AIA 304, 171 and 193 {\AA} images
(see Animation 0929-304) showing the emergence of magnetic flux and
the evolution of the filament on 2013 September 29. The upper-right
corner in panels (a)-(c) shows the enlarged image in the white
square. The blue square in panel (a) denotes the FOV of Figure 7.
The blue rectangular in panel (h) denotes the FOV of Figure 8. The
straight line `A$-$B' in panel (g) is used to obtain the stack plots
in Figures 9(a)-(b).}
              \label{Fig6}%
    \end{figure}

\begin{figure}
   \centerline{\includegraphics[width=1.0\textwidth,clip=]{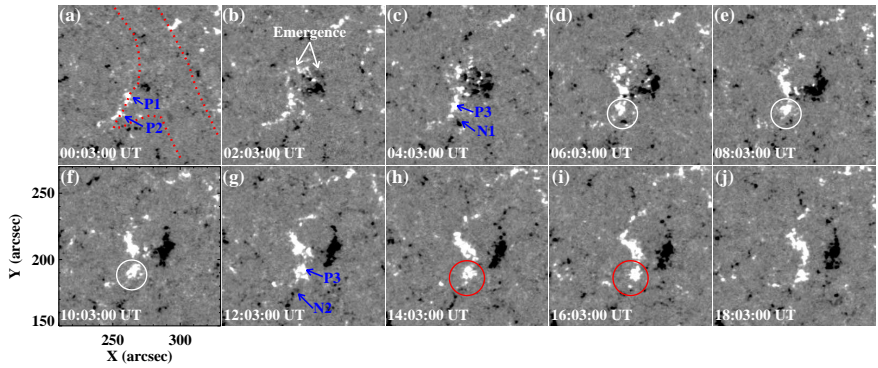}
              }
   \caption{Evolution of the magnetic structures in the vicinity of the barb in the second event. The FOV
of HMI magnetograms is represented by the blue square in Figure
6(a). The red dotted curves in panel (a) denote the filament
contours. Magnetic structures `P1' and `P2' denote the parasitic
magnetic polarities nearby the barb. Magnetic structures `P1' and
`P2' merge together into a larger one `P3'. Positive magnetic
structure `P3' cancels with negative magnetic structures `N1'
(outlined by white circles) and `N2' (outlined by red circles).}
              \label{Fig7}%
    \end{figure}

    \begin{figure}
   \centerline{\includegraphics[width=1.0\textwidth,clip=]{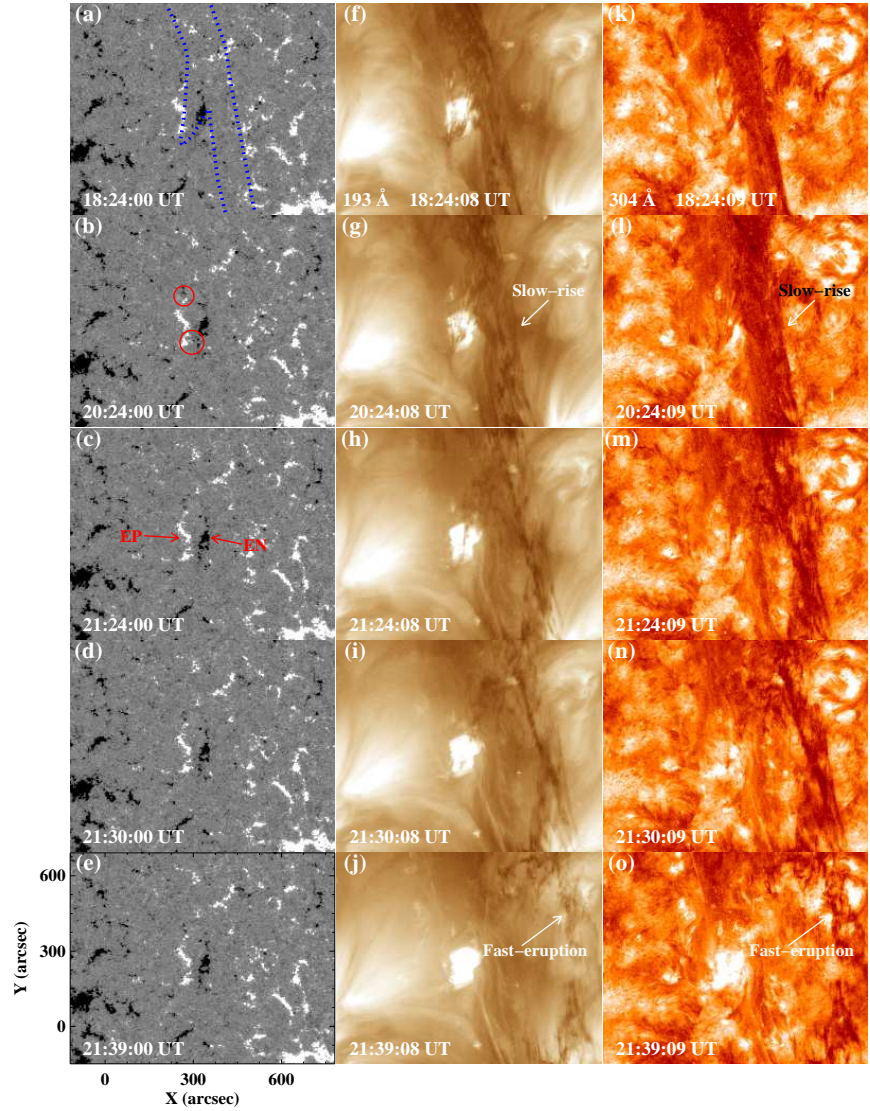}
              }
   \caption{Evolution of the emerging flux and the barb prior and during the eruption of the filament. The FOV of HMI magnetograms,
{\emph{SDO}}$/$AIA 193 and 304 {\AA} images is represented by the
blue rectangular in Figure 6(h). The blue dotted curves in panel (a)
denote the filament contours. Magnetic structures `EP' and `EN'
denote the emerging positive and negative magnetic fields. Red
circles outline the cancellation of the emerging magnetic flux (`EN'
and `EP') with nearby magnetic structures.}
              \label{Fig8}%
    \end{figure}

 \begin{figure}
   \centerline{\includegraphics[width=1.0\textwidth,clip=]{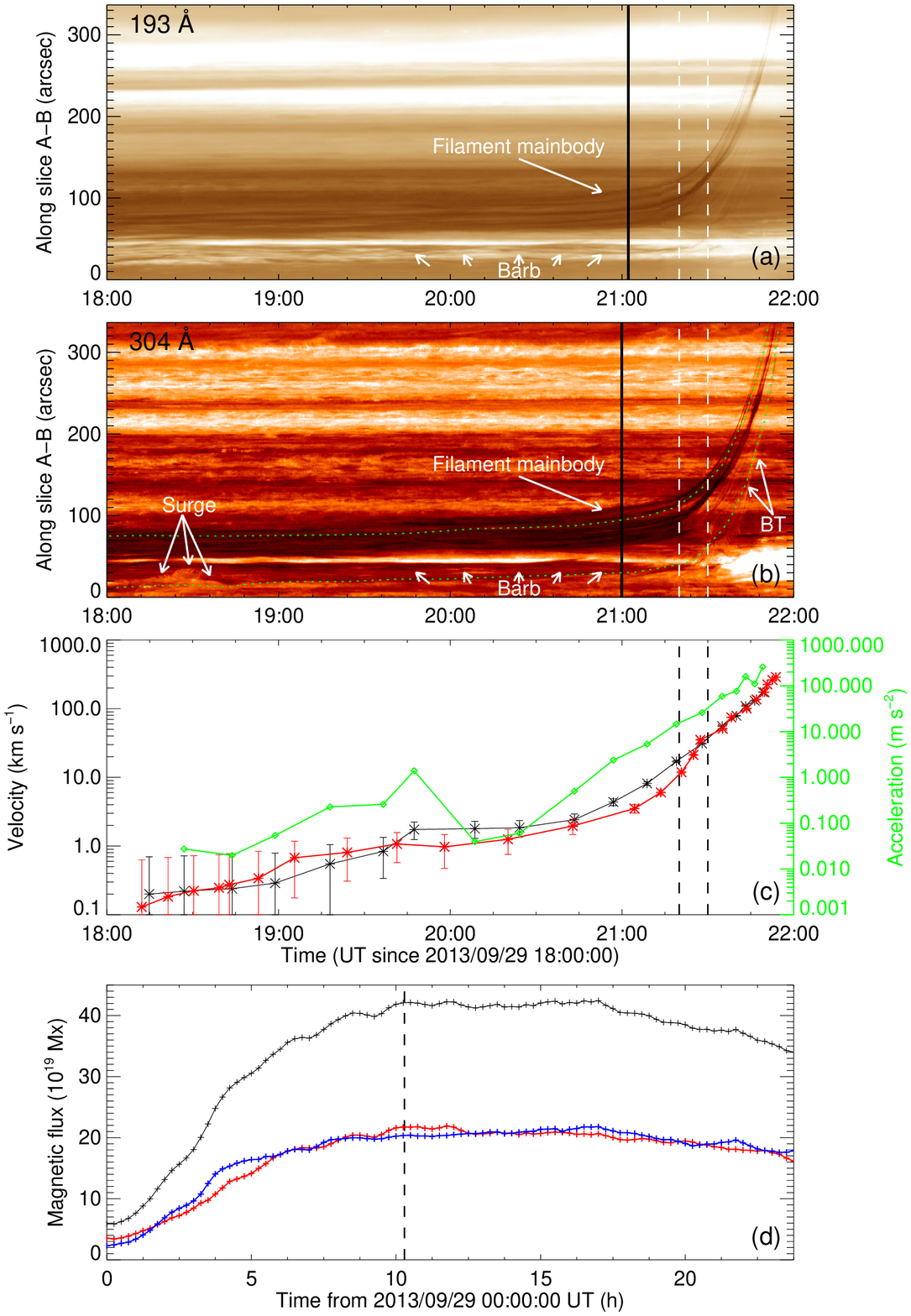}
              }
   \caption{Panel
(a)-(b): stack plots along slice `A$-$B' (blue line in Figure 6(g))
showing the evolution of the filament at AIA 193 and 304 {\AA}. Two
green dotted curves in panel (b) denote two threads of the filament
main body and the barb that are used for velocity measurements in
panel (c). `BT' denotes the bright threads during the eruption
process. Panel (c): velocity-time (black curve), acceleration-time
profiles (green curve) of a single thread of the filament main body
based on panel (b), and the velocity of the barb threads (red
curve). Dashed lines in panels (a)-(c) denote the transition time
between the slow-rise and fast-rise phases. Panel (d): temporal
variations of the unsigned positive (red curve), negative (blue
curve), and total (black curve) emerging magnetic flux. Dashed line
in panel (d) denotes the time when the total unsigned magnetic flux
reached its maximum.}
              \label{Fig9}%
   \end{figure}

At the onset of flux emergence, multiple pairs of small magnetic
patches with opposite polarities dispersedly appeared and the
positive and negative magnetic fields were interwoven together
(Figures 7(a)-(c)). Then the small patches with the same polarity
converged and merged into a large and concentrated magnetic
structure (Figure 7(d)). The emerging magnetic flux could be
identified as a dipolar region after 08:00 UT (Figure 7(e)). In the
vicinity of the barb's end, small patches of parasitic polarity were
observed (denoted by arrows `P1' and `P2' in panel (a)). The flux
emergence drove the converging motion of magnetic structure `P2'
toward `P1,' which merged together into a large magnetic structure
`P3' (panel (c)). Afterward, the positive magnetic structure `P3'
constantly canceled with nearby negative network fields (such as
magnetic structures `N1' and `N2' in panels (c) and (g)). In 193 and
304 {\AA} images, the material of the barb appeared as multiple
thread-like structures (Figure 8). Meanwhile, bi-directional mass
flows could also be observed along the axis of the barb. At 20:24
UT, the barb was elongated and the filament started to rise up
slowly (Figures 8(g) and (l)). The flux cancellation of the emerging
magnetic flux (`EN' and `EP') with nearby magnetic structures
constantly occurred (red circles in Figure 8(b)). After 21:20 UT,
the filament was accelerated and erupted rapidly.

In order to investigate the kinematic evolution of the filament in
detail, we obtained the stack plots at 193 and 304 {\AA} (Figures
9(a)-(b)) along slice `A$-$B' (Figure 6(g)). Both the filament main
body and its barb could be clearly observed in the stack plots. At
about 18:40 UT, a surge seemed to appear in the vicinity of the barb
(Figure 9(b)). This surge may be related to the later filament
eruption. By tracing a single thread of the filament main body in
the stack plot at 304 {\AA}, velocity$-$time and acceleration$-$time
profiles of the filament main body were obtained (black and green
curves in Figure 9(c)). Since about 20:30 UT, the filament started
to rise up with a small velocity less than 3 km s$^{-1}$. From 21:00
UT, the velocity showed an obvious increase. At 21:20 UT, the
velocity reached about 17 km s$^{-1}$ with the acceleration of 15 m
s$^{-2}$. Due to the uncertainty of thread location (about 3 pixel),
the uncertainty of the velocity was estimated about 1.0 km s$^{-1}$.
The velocity of the barb thread is shown by the red curve in panel
(c). In the early phase, the velocities of the two threads in the
barb and the filament main body are nearly the same and their
difference is within the error of the velocity. From about 20:45 UT,
the velocity of the barb thread is lower than that of the thread in
the filament main body. This indicates that the filament main body
drags the barb along with it. Until 21:30 UT, the barb seemed to be
divided into two parts, with the lower part remaining stable and the
top part suddenly rising up (the time interval between two dashed
lines in Figures 9(a)-(b); Figures 8(i) and (n)). The rising barb
threads were brightened during the eruption process (Figure 9(b)),
and these bright threads were probably heated by magnetic
reconnection occurring underlying the filament. The velocity of the
threads in the filament main body and the velocity of the top barb
are approximately equal to each other after 21:30 UT. Then the
filament main body erupted rapidly and its velocity increased to 170
km s$^{-1}$ (the projected velocity; the corresponding vertical
velocity is about 340 km s$^{-1}$) and the acceleration reached
about 260 m s$^{-2}$ at 21:48 UT. About 30 min later (22:12 UT),
LASCO C2 observed a CME with an angular width of 160 degree and an
average speed of 680 km s$^{-1}$. The filament eruption was also
associated with a C1.2 flare that started at 21:43 UT.

As seen from Figure 9(d), the magnetic flux of new emerging bipole
is calculated. The emerging flux in the filament channel of the
second event is about one order of magnitude less than that of the
first event. At about 10:00 UT, the total magnetic flux increased to
a maximum of 4.2$\times$10$^{20}$ Mx. The values of positive and
negative magnetic flux are approximately balanced (red and blue
curves). The average flux emergence rate is nearly
5.2$\times$10$^{15}$ Mx s$^{-1}$, about one fourth of the first
event. Until about 17:00 UT, the magnetic flux maintained at a
relatively steady value. Then the emerged magnetic flux showed an
obvious decrease due to the constant cancellation with nearby
opposite magnetic fields. By comparing the kinematic evolution of
the filament with the temporal variations of emerging magnetic flux,
we noticed that the eruption of the filament occurred in the
decaying phase of the emerging magnetic flux, with a time delay of
more than 20 hr from the start of the emergence.

\section{Summary and Discussion}

We make a comparative analysis for two filaments that showed quite
different activation in response to the flux emergence within the
filament channels. In the first event, an AR emerged below the
filament with a maximum magnetic flux of about 6.4$\times$10$^{21}$
Mx. The emerging flux resulted in the separation of the filament
into two parts, with the south end of the north part located nearby
the emerged positive polarity fields and the north end of the south
part in the emerged negative ones. Due to the interaction of the
emerging flux with the filament, bright fine-scale threads within
the filament channel extending from the AR towards two opposite
directions are clearly observed (Figure 3). This is the firstly
reported example that the filament is separated into two parts due
to an emerging AR within the filament channel. In H$\alpha$ images,
the filament is fragmented and the connectivity could not be clearly
observed. Comparison of the AIA 304 {\AA} images with H$\alpha$
images (Figures 1 and 2) shows that the chromospheric configuration
of the filament could only be observed if the optical depth in
H$\alpha$ is large enough. A similar example was presented in Li \&
Zhang (2013a), in which the spine of the filament can only be
detected in 304 {\AA} images.

For the first event, the filament did not erupt within 3 days,
instead, it was separated into two parts (north and south
filaments). The close neighborhood and common direction of the
bright threads in the filament and the open AR fan loops suggest
similar magnetic connectivity of these two flux systems. These
threads in both parts of the filament may imply the configuration of
the magnetic field lines (Li \& Zhang 2013b; Joshi et al. 2014). On
the 3rd and 4th day after the emergence of the AR , the AR started
to decay. Near that time, the emergence and cancellation of other
small magnetic polarities occurred nearby the barbs of the southern
filament. These small-scale magnetic fields seemed to disturb the
stability of the south filament, which resulted in the partial
eruption of the filament. The statistics showed that the shorter
distance between the new active regions and quiescent filaments, the
less time between active region appearance and filament
disappearance (Bruzek 1952; Balasubramaniam et al. 2011). Thus the
time interval of 4 days is too long and it is not reasonable to
correlate the AR emergence within the filament channel with the
later eruption. A case of filament eruption on the 4th day after a
new flux emergence away from the filament channel was analyzed by
Feynman \& Martin (1995). The active region continued to emerge on
the day of the eruption and they attributed the eruption to the flux
emergence. However, in our first event, the AR emergence ceased
about 2 days before the eruption. We suggest that the eruption of
the south filament is caused by the emergence and cancellation of
the small-scale magnetic flux (Figure 5) nearby the barbs about 20
hr before its eruption.

In the second event, the emerging flux occurred nearby a barb of the
filament with a maximum magnetic flux of 4.2$\times$10$^{20}$ Mx,
about one order of magnitude less than that of the first event. The
flux emergence probably caused the global destabilization of the
filament and resulted in the final eruption of the entire filament
about 20 hr after the start of flux emergence. By investigating the
HMI magnetograms, it was found that the emerging flux drove the
convergence of two patches of parasitic polarity in the vicinity of
the barb. The cancellation between the parasitic polarity and nearby
network fields was also observed. The evolution (flux convergence
and cancellation) of parasitic polarity nearby the barb probably
affected the stability of the filament. The velocity of the thread
in the filament main body is larger than that of the barb thread.
This indicates that the global destabilization of the filament is
triggered and the filament main body drags the barb along with it.
However, the former case (barb pushing the main body) would be a
strong evidence for flux near the barb leading to destabilization of
the filament. The main body of the filament in the second event
underwent a transition from slow-rise phase to fast-rise phase. The
velocity of the filament increased to 170 km s$^{-1}$ and the
acceleration reached about 260 m s$^{-2}$ about 20 min after the
start of the fast-rise phase.

According to the models of Aulanier et al. (1998, 1999) and
Dud{\'{\i}}k et al. (2008), the plasma of barbs are supported by
magnetic dips that are formed due to the presence of parasitic
polarities in the filament channel. The change of the parasitic
patches nearby the barbs is responsible for the evolution of barbs.
In our study, the observed flux convergence and cancellation of
parasitic polarity disturbed the magnetic fields passing through the
barb, and the tether-cutting reconnection (Moore et al. 2001)
between the field lines supporting the barb material (magnetic dips
nearby the barb) and nearby network fields might be triggered. Then
the stability of the filament was catastrophically destroyed and the
filament was erupted rapidly.

The relation of flux emergence and the filament eruption has been
analyzed in previous studies (Sun et al. 2012; Mandrini et al.
2014). Joshi et al. (2011) found the coincidence of the X-ray
precursor phase with the flux emergence at the flaring location, and
suggested that the flux emergence and pre-flare activities played a
crucial role in destabilizing the filament. In our work, the AR in
the first event emerged within the filament channel and the filament
did not erupt within 3 days. No barbs of the filament are observed
nearby the emerging AR. Afterwards, the emergence and cancellation
of small-scale magnetic flux occurred nearby the barbs and the south
filament erupted. For the second event, the emerging magnetic flux
nearby the barb followed by flux cancellation resulted in the
filament eruption about 20 hr later despite the fact that the
emerging flux was one order of magnitude less than that of the AR in
the first event. Our findings imply that the location of emerging
flux within the filament channel is probably crucial to filament
evolution. If the flux emergence appears nearby the barbs, flux
cancellation of emerging flux with the filament magnetic fields is
prone to occur. The flux cancellation results in the tethers to cut
and reconnection to take place beneath the filament (Vemareddy et
al. 2012; Kuridze et al. 2013; Huang et al. 2014; Jing et al. 2014).
Thus flux emergence/cancelation and tether-cutting reconnection
around the barbs of the filament cause the destabilization of the
filament and the onset of its eruption. If the flux emergence within
the filament channel is far away from the barbs, the occurrence of
flux cancellation between emerging flux and the filament magnetic
fields is not easy, and the equilibrium of filament is not
destroyed. The comparison of the two events shows that the emergence
of an entire AR may still not be enough to disrupt the stability of
a filament system and the actual eruption does occur only after the
flux cancellation sets in. Zuccarello et al. (2014) observed flux
cancellation toward the compact polarity inversion line during the
8-10 hr preceding the eruption. Sterling et al. (2010) also
suggested that the main filament ejection was triggered by flux
cancelation between the positive flux elements and the surrounding
negative field. In our work, the filament eruptions studied are
associated with flux emergence, but occur only after completion of
the emergence and/or after the onset of the magnetic cancellation
following the emergence. Although we do not perform modelling of the
magnetic field of the filaments and their barbs, our observations of
the emergence followed by cancellation in relation to the timing of
filament eruptions add to the debate on the mechanism behind
destabilization of solar structures. More samples need to be
thoroughly analyzed to obtain a statistic about the relationships
between filament evolution and flux emergence and cancellation in
the filament environment.

\begin{acks}
We acknowledge the \emph{SDO}/AIA, HMI and NSO$-$GONG for providing
data. This work is supported by the National Basic Research Program
of China under grant 2011CB811403, the National Natural Science
Foundations of China (11303050, 11025315, 11221063 and 11003026),
the CAS Project KJCX2-EW-T07 and the Strategic Priority Research
Program$-$The Emergence of Cosmological Structures of the Chinese
Academy of Sciences, Grant No. XDB09000000. H. S. Ji was supported
by NSFC grants 11333009 and 1117306.
\end{acks}

\end{article}

\end{document}